# A Fast Algorithm for Line Clipping by Convex Polyhedron in $E^3$

Václav Skala[1]
Department of Informatics and Computer Science
University of West Bohemia
Univerzitní 22, Box 314, 306 14 Plzen
Czech Republic
e-mail: skala@kiv.zcu.cz http://herakles.zcu.cz/~skala

## Abstract

A new algorithm for line clipping against convex polyhedron is given. The suggested algorithm is faster for higher number of facets of the given polyhedron than the traditional Cyrus-Beck's and others algorithms with complexity $O(N)$. The suggested algorithm has $O(N)$ complexity in the worst case and expected $O(\sqrt{N})$ complexity. The speed up is achieved because of "known order" of triangles. Some principal results of comparisons of selected algorithms are presented and give some imagination how the proposed algorithm could be used effectively.

## 1. Introduction

A problem of line clipping against convex polyhedron in $E^3$ can be solved by Cyrus-Beck's (CB) algorithm [CYR79a] for three dimensional case. Many algorithms for line clipping in $E^2$ and $E^3$ have been published so far, with $O(N)$ or $O(\lg N)$ complexities. The $O(\lg N)$ complexity in $E^2$ is achieved because of convexity feature of the given polygon and the **known order** of vertices or edges. For comparisons and references see [SKA93a], [SKA94a]. If preprocessing is used the processing complexity can be decreased to O(1), see [SKA96b], [SKA96c].

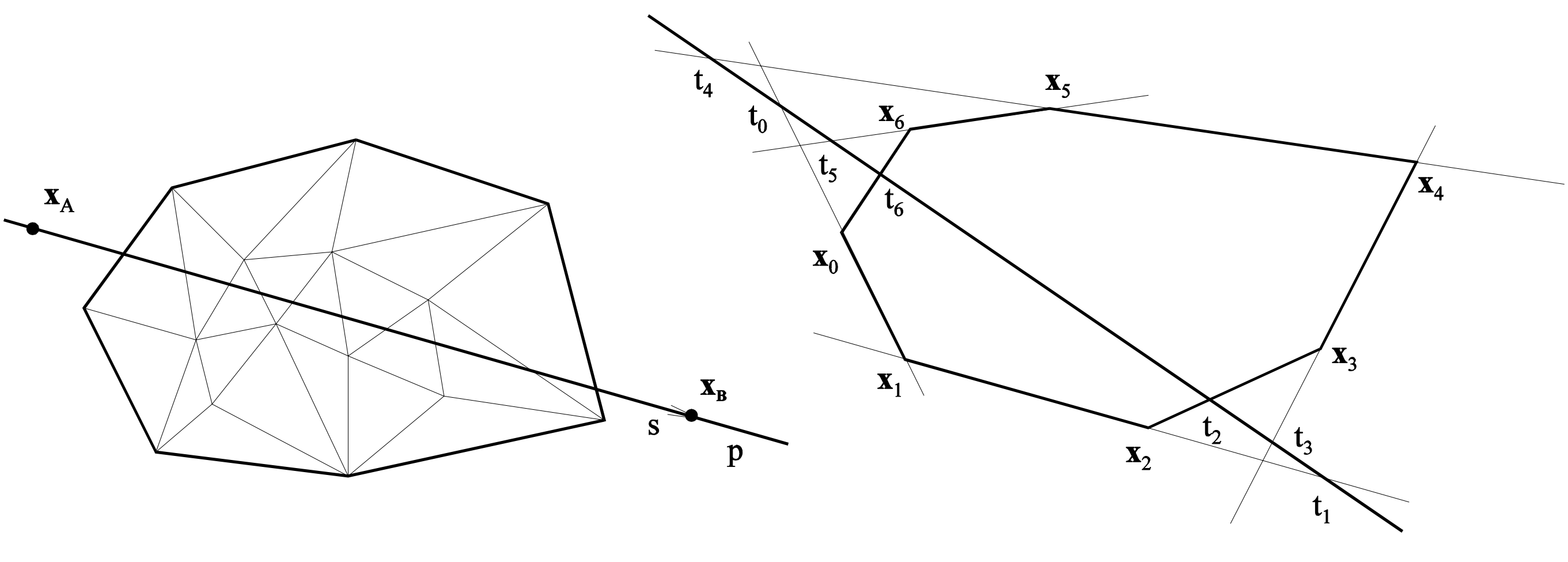

Line clipping by convex polyhedron in $E^3$

Figure 1

[1] Supported by the grant UWB-156/1995-6

Nevertheless algorithms for $E^3$ case are mostly based on the CB algorithm modified for $E^3$ and restricted to convex polyhedra, to orthogonal or pyramidal volumes, see [FOL90a], or based on direct intersection computation of a polyhedron facet (triangle) and the clipped line *p*, see fig.1. Because the line clipping against a convex polyhedron in $E^3$ is a bottleneck of many applications it would be desirable to use the fastest algorithm for clipping even it is of complexity $O(N)$. It is necessary to point out that a number of facets *N* is expected to be high especially if sphere or cylindrical volumes are approximated by polyhedra.

## 2. Cyrus - Beck's algorithm

The main disadvantage of the CB algorithm is a direct line intersection computation for all planes which form the boundary of the given convex polyhedron. It means that ***N - 2* of intersection computations are wasted** if *N* is a number of facets of the given convex polyhedron. That is a very substantial because the average number of facets of the given polyhedron is high (a number of facets might easily reach for a sphere approximation value $10^4$).

The efficiency of the CB algorithm is determined by a simple algorithm for direct intersection computation of a line with a plane in $E^3$. It is obvious that with growing number of facets of the given polyhedron the efficiency of the CB algorithm decreases as many invalid intersection points are computed, see alg.1 for shorten version of the CB algorithm.

```
procedure Clip_3D_CB ( xA , xB );
begin { ni is a normal vector of the i-th facet       }
    { and ni must point out of the convex polyhedra }
    { !! all normal vectors ni are precomputed !!   }
  tmin := 0.0;    tmax := 1.0;     s :=  xB - xA;
  { for the line clipping tmin:= -∞; tmax := ∞;  }
  i := 1 ;
  for  i:= 1 to n do        { N is a number of facets }
  begin
        ξ := s^T ni;       si := xi - xA;
        if  ξ <> 0.0 then
        begin t := si^T ni / ξ;
                if ξ > 0.0 then tmax := min ( t , tmax)
                          else tmin := max ( t , tmin)
        end
        else Special case solution;{ line is parallel to a facet }
        i := i + 1
  end;
  if tmin > tmax then EXIT; { !! < tmin, tmax > = ∅ }
 { recompute end-points of the line segment if changed }
 { for lines points xA , xB must be always recomputed  }
  if tmax < 1.0 then xB := xA + s tmax;
  if tmin > 0.0 then xA := xA + s tmin;
  SHOW_LINE ( xA , xB );
end { Clip_3D_CB };
```

Algorithm 1

Let us consider only triangular facets of the given convex polyhedron in the following (generally it is not necessary). For the given line *p* (line *p* varies!) it is necessary to find an **effective test** whether the line *p* intersects the triangle, see fig.2.

The intersection points of the line and the triangular facet can be directly computed as a solution of the following linear parametric equations

$$\mathbf{x}(t) = \mathbf{x_A} + \mathbf{s}\,t \qquad t \in (-\infty, \infty) \qquad (\alpha)$$

$$\mathbf{x}(p,q) = \mathbf{x_O} + \mathbf{s_1}\,p + \mathbf{s_2}\,q \qquad p, q \in \langle 0, 1 \rangle \;\&\; p + q \leq 1 \qquad (\beta)$$

e.g. in the matrix form

$$\left[\, \mathbf{s_1} \mid \mathbf{s_2} \mid -\mathbf{s} \,\right] \cdot \begin{bmatrix} p \\ q \\ t \end{bmatrix} = \mathbf{x_A} - \mathbf{x_O}$$

where $\alpha$ is an equation for a given line,

$\beta$ is an equation for a given triangular facet, see fig.2.

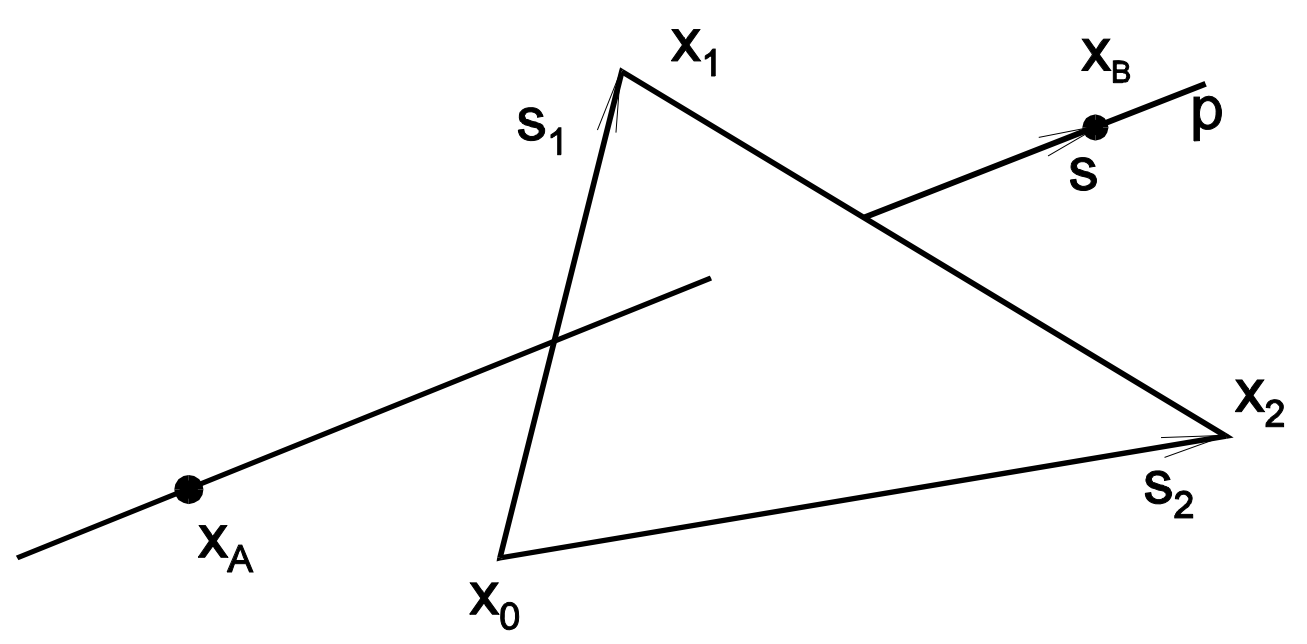

Line p intersects the triangular facet

Figure 2

Because all direct tests are nearly as complex as computations made in the CB algorithm it is necessary to find an effective method for selection of facets that might be intersected by a line *p*.

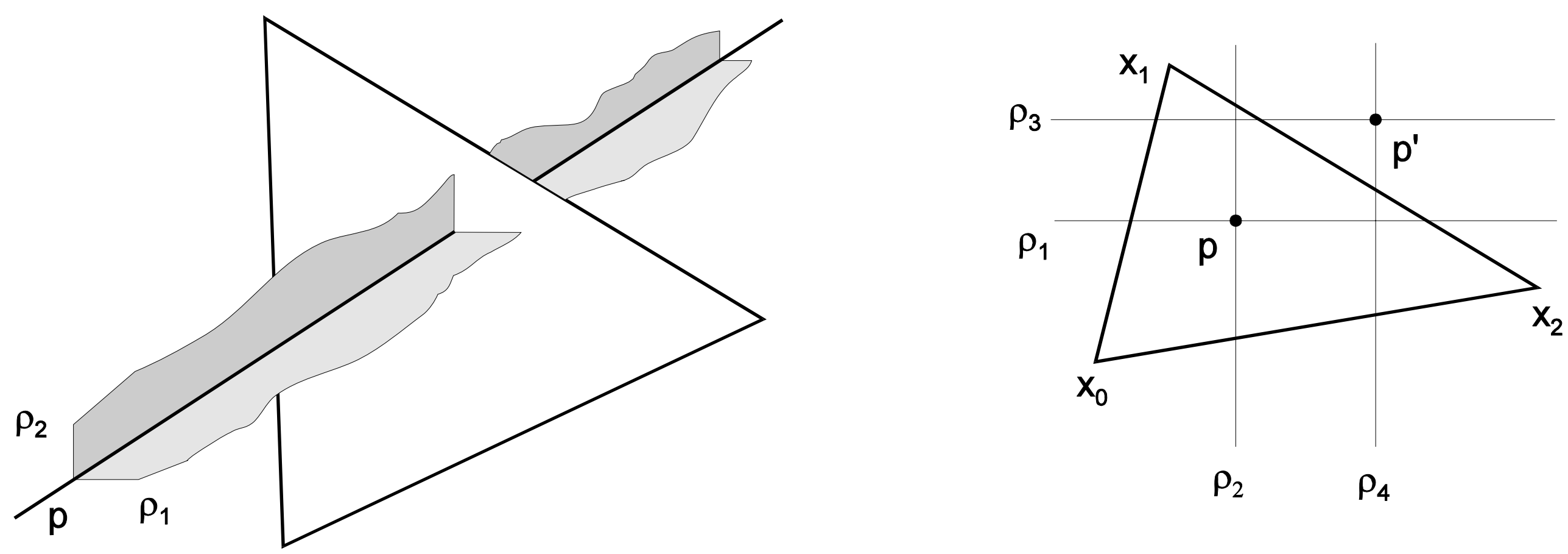

Usage of two planes for line definition

Figure 3

### 3. Algorithm based on planes

This algorithm [SKA96a] is based on the idea that a line p can be defined as an intersection of two non-collinear planes $\rho_1$ and $\rho_2$, see alg.2. It can be seen that if the line *p* intersects the given triangle then planes $\rho_1$ and $\rho_2$ intersect the given triangle, too, but not vice versa, see fig.3. The line *p'* that is defined as the intersection of $\rho_3$ and $\rho_4$ planes does not intersect the triangle.

It is possible to test all triangles (facets) of the given polyhedra against $\rho_1$ and $\rho_2$ planes. If both planes $\rho_1$ and $\rho_2$ intersect the given triangle (facet) then compute the detailed intersection test. The intersection of the given plane $\rho_i$ and the triangle exists **if and only if** two vertices $\mathbf{x_A}$ and $\mathbf{x_B}$ of the triangle exist so that

$$\text{sign}(F_i(\mathbf{x_A})) \neq \text{sign}(F_i(\mathbf{x_B}))$$

where $F_i(\mathbf{x}) = A_i\, x + B_i\, y + C_i\, z + D_i$ is an equation for the i-th plane $\rho_i$, i=1,2.

The substantial advantage is that $\rho_1$ and $\rho_2$ planes can be taken as parallel with any coordinate axes. Those planes are usually called „diagonal". In that case the functions $F_i(\mathbf{x})$ can be simplified so that

$F_1(\mathbf{x}) = A_1\, x + C_1\, z + D_1$ for plane $\rho_1$
$F_2(\mathbf{x}) = B_2\, y + C_2\, z + D_2$ for plane $\rho_2$

It is possible to divide $F_1(\mathbf{x})$ by $A_1 \neq 0$. Using this approach we save one addition and two multiplication, similarly for $F_2(\mathbf{x})$, per facet.

Unfortunately there is some principal inefficiency in this proposed solution as the separation function $F_1(\mathbf{x})$, resp. $F_2(\mathbf{x})$ are computed more times than needed because **every vertex is shared** by many triangles. Therefore it is convenient if the values $\text{sign}(F_1({}^i\mathbf{x}_k))$ are precomputed ( ${}^i\mathbf{x}_k$ is the k-th vertex of the i-th facet) and stored in a separate vector, see alg 3. This modification significantly improves the efficiency of the algorithm. It is possible to select planes $\rho_1$ and $\rho_2$ as two „diagonal " planes in order to avoid singular cases, see [SKA96a] for detail description and comparison. This algorithm is still of *O(N)* complexity.

**procedure** CLIP_3D_MOD ( $\mathbf{x_A}$ , $\mathbf{x_B}$ );
**begin**
  $t_{min}$ := 0.0;   $t_{max}$ := 1.0;   i := 1;   $\mathbf{s}$ := $\mathbf{x_B} - \mathbf{x_A}$;
  { for the line clipping $t_{min}$:= $-\infty$; $t_{max}$ := $\infty$; }
  { $\rho_1 : F_1(\mathbf{x}) = A_1\, x + C_1\, z + D_1 = 0$     $\rho_2 : F_2(\mathbf{x}) = B_2\, y + C_2\, z + D_2 = 0$ }
  **for** k := 1 **to** $N_v$ **do**  { $N_v$ number of vertices }
    $Q_k$ := sign($F_1(\mathbf{x}_k)$); { $Q_k$ is a vector of int or char types }
  **for** i :=1 **to** N **do**
  **begin**
    { ${}^i\mathbf{x}_k$ means a k-th vertex of the i-th triangle }
    { INDEX(i,k) gives the index of k-th vertex of the i-th triangle, i.e. ${}^i\mathbf{x}_k = \mathbf{x}_{INDEX(i,k)}$ }
    **if** $Q_{INDEX(i,0)} = Q_{INDEX(i,1)}$ **then**
      **if** $Q_{INDEX(i,0)} = Q_{INDEX(i,2)}$ **then goto** 1;
      { do nothing $\rho_1$ does not intersect the i-th triangle }
    **if** sign($F_2(\mathbf{x}_{INDEX(i,0)})$) = sign($F_2(\mathbf{x}_{INDEX(i,1)})$) **then**
      **if** sign($F_2(\mathbf{x}_{INDEX(i,0)})$) = sign($F_2(\mathbf{x}_{INDEX(i,1)})$) **then goto** 1;
    { both planes $\rho_1$, $\rho_2$ intersect the i-th triangle }
    { detailed test finds a value $t_{min}$ and $t_{max}$ using a single step of the CB algorithm }
    { ---------------------------------- }

```
        { n_i must point out of the given convex polyhedron }
        ξ := s^T n_i;        s_i := x_i - x_A;
        if  ξ <> 0.0 then
        begin t := s_i^T n_i / ξ;
                if ξ > 0.0 then t_max := min ( t , t_max)
                            else t_min := max ( t , t_min);
        end
        else Special case solution;{ line is parallel to a facet }
        { ---------------------------------- }
  1:    i := i + 1;
  end;
  { if  t_min > t_max  then no intersection point exists }
  if t_min ≤ t_max then SHOW_LINE (x(t_min), x(t_max));
end { of CLIP_3D_MOD };
```

Algorithm 2

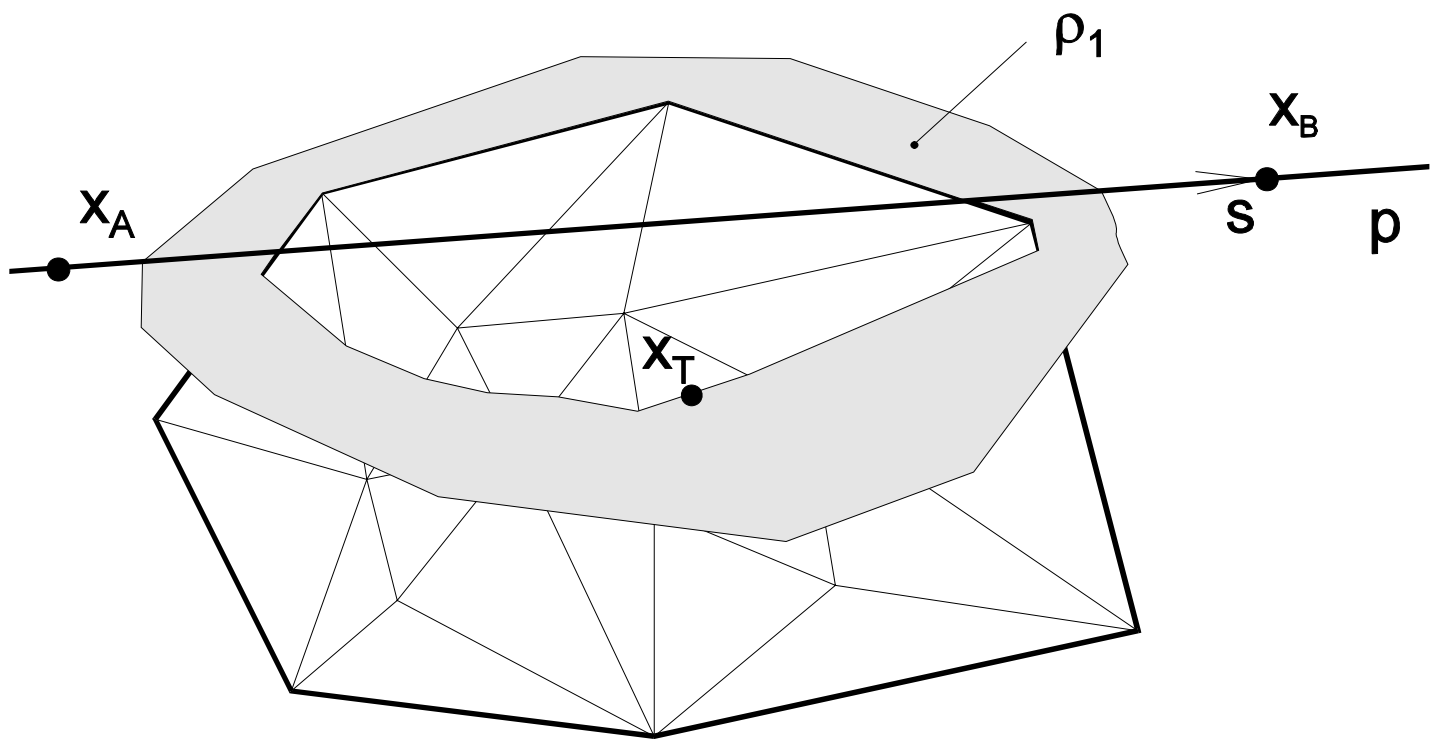

Definition of the plane $\rho_1$

Figure 4

## 4. Proposed method

Finding a facet candidate for the intersection point is a quite complex task and without knowing the "order" of facets an algorithm will be generally of *O(N)* complexity. Nevertheless we can select any triangle (facet) $\tau_k$ on which the given line does not lie (it means that the definition of the strating point of the search is *O(1)* ). If we take any point inside of the triangle $\tau_k$, i.e. a centriod of the facet $\mathbf{x}_T$, and the given line we obtain a definition of a plane $\rho_1$, see fig.4 on which the line lies. It can be seen that we could develop more efficient strategy for testing triangles if we know facets that are intersected by the plane $\rho_1$. If non-convex polyhedron is considered then two or more separate “rings of triangles“ are necessary to solve but only one will be detected and solve. Therefore we must consider convex polyhedra only. Unfortunately the situation is not as simple as in algorithm for line clipping with $O(\lg N)$ complexity in $E^2$, see [SKA94a].

In many applications, data structures that define a convex polyhedron contain information about neighbours of the given triangular facet, see fig.5. It is obvious that we can easily detect which edge of the given triangle $\tau_k$ is intersected by the plane $\rho_1$. In the next step we can take its neighbour which has a common edge with the triangle $\tau_k$, etc. Only the facets reached by these neighbour construction have to be taken into consideration, i.e. have to be tested against $\rho_2$.

Because of that "knowledge of order" we will have to test significantly less triangles than *N*, but in the worst case we will have to test all *N* facets. It means that the algorithm is of *O(N)* complexity in the worst case.

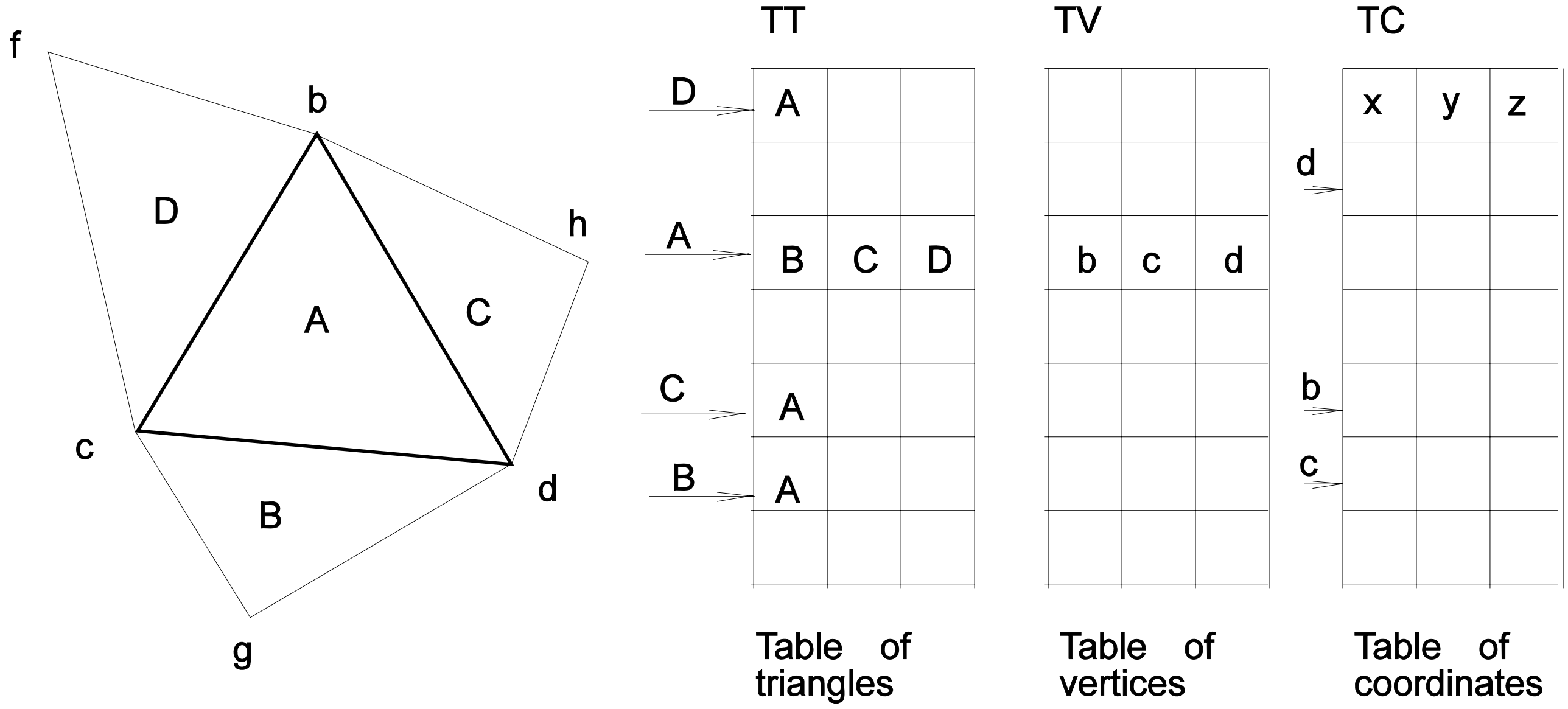

Data structure with information about neighbours
Figure 5

Let us consider a surface of a convex polyhedron, e.g a sphere approximation, that is intersected by a plane $\rho_1$. A number of intersected facets by a plane can be estimated as $\sqrt{N}$ in average. This algorithm is briefly described in alg.3 and it is necessary to point out that there are some small obstacles in solving singular cases, especially when the plane $\rho_1$ intersects the triangle in vertex etc. The presented approach can be easily modified for non-triangular facets if **get_next** procedure is modified properly.

```
procedure SQRT_CLIP ( xA , xB );
begin  tmin := 0.0;    tmax := 1.0;    s := xB - xA;    { for the line clipping tmin:= -∞; tmax := ∞; }
       COMPUTE ( xA, xB, k, ρ1 ); { finds a convenient facet τk and the plane ρ1 through xA , xB , xT }
       COMPUTE_ORTHOGONAL ( xA , xB , ρ2 );
       { computes coefficients of the orthogonal plane to ρ2 }
       k0 := -1; { set index of previous facet }
       while k ≠ k0 do
       begin { test the facet τk against the plane ρ2}
             if TEST (τk, ρ2) then
             begin  { a single step of the CB algorithm }
                    { nk must point out of the given convex polyhedron }
                    ξ := s^T nk;        sk := xk - xA;
                    if ξ <> 0.0 then
                    begin  t := sk^T nk / ξ;
                           if ξ > 0.0 then tmax := min ( t , tmax) else tmin := max ( t , tmin);
                    end
                    else Special case solution;{ line is parallel to a facet }
             end;  { if tmin > tmax then no intersection point exists }
             i := k0;              k0 := k;          k := get_next ( k , i);
             {get next facet to τk intersected by the plane ρ1;different from previous facet τk0 }
       end;
       if tmin ≤ tmax then SHOW_3D_LINE(x(tmin),x(tmax));
end { SQRT_CLIP };
```

Algorithm 3

## 5. Experimental results

The proposed algorithm has been tested against the CB algorithm, see alg.1, and against the algorithm that uses two planes, see alg.2. Data sets of two points that defines lines have been randomly and uniformly generated inside a sphere in order to eliminate an influence of rotation. Convex polyhedra were generated as *N* - sided convex polyhedra that consist of triangular facets and were inscribed into a smaller sphere.

Let us introduce coefficients of efficiency as

$$\nu_1 = \frac{T_{CB}}{T} \qquad \nu_2 = \frac{T_\rho}{T}$$

where $T_{CB}$, $T_\rho$, T are execution times in [s] needed by the CB algorithm (CLIP_3D_CB), algorithm (CLIP_3D_MOD) that uses two planes and the proposed algorithm (SQRT_CLIP).

Results obtained from experiments are shown in tab.1 and tab.2 for two fundamental cases when any line does not intersect the given polyhedra (0% of intersections) and when all lines intersect (100% of intersections) the given polyhedra and 10 000 clipped lines. All tests were made on PC 486/33 Mhz and all special cases were solved properly.

| N | 10 | 20 | 50 | 100 | 200 | 500 | 1000 | 2000 | 4000 |
|---|---|---|---|---|---|---|---|---|---|
| $T_{CB}$ | 0.9 | 2.5 | 7.8 | 16.4 | 33.0 | 83.5 | 167.5 | 335.5 | 672.1 |
| $T_\rho$ | 0.7 | 1.2 | 3.4 | 6.1 | 11.4 | 26.9 | 52.6 | 104.7 | 306.7 |
| T | 1.9 | 2.8 | 5.8 | 7.4 | 11.4 | 16.9 | 25.8 | 36.8 | 54.9 |
| $\nu_1$ | 0.5 | 0.9 | 1.3 | 2.2 | 2.8 | 4.9 | 6.5 | 9.1 | 12.2 |
| $\nu_2$ | 0.4 | 0.4 | 0.6 | 0.8 | 1.0 | 1.6 | 2.0 | 2.9 | 5.6 |

Efficiency coefficients ν for 0% of intersections

Table 1

| N | 10 | 20 | 50 | 100 | 200 | 500 | 1000 | 2000 | 4000 |
|---|---|---|---|---|---|---|---|---|---|
| $T_{CB}$ | 1.3 | 3.1 | 8.1 | 16.6 | 33.4 | 84.5 | 169.1 | 338.0 | 676.5 |
| $T_\rho$ | 4.4 | 5.4 | 8.1 | 10.6 | 16.2 | 31.6 | 54.8 | 101.0 | 190.7 |
| T | 5.4 | 6.4 | 9.2 | 10.1 | 13.9 | 18.6 | 25.8 | 34.0 | 49.1 |
| $\nu_1$ | 0.2 | 0.5 | 1.0 | 1.6 | 2.4 | 4.5 | 6.5 | 10.0 | 13.8 |
| $\nu_2$ | 0.8 | 0.8 | 0.9 | 1.0 | 1.2 | 1.7 | 2.1 | 2.9 | 3.9 |

Efficiency coefficients ν for 100% of intersections

Table 2

The proposed SQRT_CLIP algorithm is always faster than the CB algorithm for $N \geq 50$, see tab.1 and tab.2, and for $N \geq 200$ is faster than the algorithm that utilizes planes, see alg.2 and [SKA96a].

## 6. Conclusion

The new efficient algorithm for clipping lines against convex polyhedron was developed with $O(N)$ complexity in the worst case and expected $O(\sqrt{N})$ complexity. Algorithm does not strictly require triangular facets and can be easily modified. The given facets should be oriented. All tests were implemented in C++ on PC 486/33 MHz.

## 7. Acknowledgments

The author would like to express his thanks to students of Computer Graphics courses at the University of West Bohemia in Plzen and Charles's University in Prague who stimulated this work, especially to P.Sebránek and J.Jirák for their careful tests verification and implementation of the proposed algorithm. Anonymous referees helped a lot as their suggestions and critical comments improved the manuscript significantly.